# A theory of viscoelastic nematodynamics

A. I. Leonov[a][1] and V.S. Volkov[b]

[a] *Department of Polymer Engineering, The University of Akron,*

*Akron, Ohio 44325-0301, USA.*

[b] *Laboratory of Rheology, Institute of Petrochemical Synthesis, Russian*

*Academy of Sciences, Leninsky Pr., 29, Moscow 117912 Russia.*

**Abctract**

A nonlinear viscoelastic theory of nematodynamic type is developed for nematic liquid crystalline (LC) semi-flexible polymers. A measure of transient elastic strain due to the change in length of macromolecular strands under stress, and the director of unit length are employed in the theory as hidden variables. In the marked contrast to the common theoretical approaches to low molecular nematics, the effect of director's space gradient is neglected in the present theory. Nevertheless, the theory allows describing nonlinear anisotropic viscoelasticity and evolution equation for the director in flows of LC polymers. When LC macromolecules are relatively rigid or when they are soft but the flow is slow, a weakly nonlinear viscoelastic anisotropic behavior is described by few temperature dependent parameters. In the infinitesimal case the evolution equation for director reminds the Ericksen's equation, but with an additional relaxation term. The present theory can also be applied for analyzing flows of concentrated polymer suspensions and nano-composites filled with uniaxially symmetric particles.

## 1. Introduction

Polymer liquid crystals consisting of relatively rigid macromolecules were discovered relatively recently [1]. They belong to a class of anisotropic polymer materials that combine the properties of liquid crystals and liquid or solid polymers. There are two basic types of LC polymeric liquids: (i) lyotropic and thermotropic LC polymers

---

[1] Corresponding author. Fax: +011-330-258-2339

*E-mail address*: leonov@uakron.edu (A.I. Leonov)

consisting of rigid enough monomeric units, and (ii) with flexible spacers where the rigid monomeric units belong either to backbone or to side chains. Additionally, there is also a class of solid LC elastomers where the LC macromolecules in elastomeric liquid of the second type are crosslinked. It should also be mentioned that the LC polymers of the first type are the most industrially developed, although these of the second type have the most perspective properties and applications. In contrast to the common polymeric fluids, LC polymers exhibit during flow anisotropic viscoelastic properties characterized by both the anisotropic relaxation times and viscosities. The solid-like LC elastomers also exibit nonlinear anisotropic mechanical and other properties in thermodynamic equilibrium, in addition to nonlinear anisotropic relaxation properties characteristic for LC polymer fluids.

A lot of theoretical effort was made to develop molecular theories modelling the lyotropic LC polymers. Those are the dominant mainstream theories today with numerous publications [2-6]. These theories typically use the Doi molecular long rigid rod approach [2-4] with further elaborations. Also, the Poisson-Bracket continuum approach related to irreversible thermodynamics was additionally elaborated [7,8] to develop constitutive equations for LC polymers. This theory reduces to the Doi theory in the homogeneous (monodomain) limit. It should be noted that Dzyaloshinskii and Volovick [9] were first who developed the Poisson-Bracket formulation of nonlinear dynamics low molecular weight (LMW) liquid crystals. There are also the attempts to use the Leslie-Ericksen theory [10], elaborated for LMW liquid crystals in the description of thermotropic LC polymers. All the above theories [2-10] employ the same state variables as in the case of LMW liquid crystals, i.e. the director $\underline{n}$ (or the second rank order tensor), and the director's space gradient, $\nabla \underline{n}$. We should also note that none of the abovementioned theories for LC polymers is capable to describe the viscoelastic experimental data for LC polymers, even for such a fundamental characteristics as the dynamic modulus. E.g. in the homogeneous (monodomain) limit they incorrectly predict that the dynamic modulus vanishes when the director is parallel to either the flow or gradient direction (see e.g. [11]).

Unlike the LMW liquid crystals, where the common solid like elasticity is insignificant and the rotation mobility of small molecules, idealized as rigid rods, dominates, in the high molecular polymeric liquid crystals with a typical degree of polymerization of several hundreds, the new quality due to the long chain molecules nature comes into the action. This is the flexibility of long molecular chains even with highly rigid monomer units, not mentioning the LC polymer with flexible spacers and LC elastomers. The molecular nature of this flexibility might be different for the two above types of LC polymers.

A good example where the molecular flexibility plays a dominant role is the equilibrium behavior of LC elastomers. De Gennes [12] was the first who using a continuum thermodynamic approach analysed the weakly elastic case involving the director space gradient as a thermodynamic variable. De Gennes' approach was later developed without director space gradient for finite deformations and finite rotations in many Warner papers (e.g. see recent publication [13] and references there), where the statistical theory of rubber elasticity was extended for the case of nematic type anisotropy.

Involving the molecular flexibility in the flows of LC polymers of both types was first made in original Russian versions of papers published later in English [14-16]. In these papers, a continuum theory of anisotropic weak viscoelasticity of LC semi-flexible polymers has been initiated, without thermodynamic analysis. The theory [14-16], extended the Maxwell constitutive equation to an anisotropic liquid with uniaxial symmetry and small transient elastic deformations, was aimed at studying anisotropic viscoelasticity of monodisperse LC polymers in slow relaxation region. This approach demonstrated a good comparison of the model predictions with data by Porter and Johnson [17].

The description of the irreversible phenomena starts with the choice of the state variables and the free energy function. In minimal theories they usually mimic the theories developed for the equilibrium case. For LC semi-flexible polymers (i.e. those with flexible spacers), these natural variables are the director that characterizes uniaxial anisotropy, transient elastic strains, and related to anisotropy internal rotations. Pleiner and Brand [18,19] were the first who involved this approach, however not in the minimal sense. They added the space gradients of thermodynamic state variations in the theoretical description of LC polymers with flexible spacers and used the linear thermodynamic scheme with account of internal rotations. The paper [19] demonstrated that the internal rotations might essentially contribute to the interpretations of fast effects, especially for the sound propagation in LC polymers with side chains. Rey [20,21] applied the thermodynamic scheme for the description of some nonlinear phenomena in flow of LC polymers with flexible spacers. He employed as the state variables the director and transient elastic strains. However his approach was restricted to small transient strains with very specific terms in free energy of doubtful significance, and contained several fundamental errors. The most important error is the occurrence of asymmetrical stress component in his thermodynamic scheme. This approach also leads to incorrect result that the dynamic moduli for LC polymers, vanishes when the director is parallel to the flow or gradient direction [21].

At present, a thermodynamically related continuum theory of viscoelasticity for LC polymers, so called "polymer nematodynamics" is not much developed. The recent attempt [22] to develop a molecular approach to the polymer nematodynamics has not produced a closed set of anisotropic viscoelastic constitutive relations, even in the linear case. It should also be noted that while a comprehensive microscopic theory for low molecular weight LC's does not currently exist, the dynamic properties of these LC's have been successfully described by continuum theories.

The present paper develops a thermodynamically related continuum theory of anisotropic viscoelasticity ("polymer nematodynamics") for nematic LC polymers with extendable macromolecules. The main objective here is deriving nonlinear constitutive equations, which describe the viscoelastic and orientation properties of polymer nematics within the framework of non-equilibrium thermodynamics. This approach has demonstrated its power in formulation of constitutive equations for polymers with flexible chains [23,24].

The most important feature, which the present theory attempts to describe, is the anisotropic viscoelasticity of LC polymers, affected in flow by their orientational elasticity. This complicated behavior is the result of anisotropic properties and interactions of LC macromolecules, along with their partial flexibility. It is well known that elastic strains in LMW nematics mostly occur due to the orientation deformation, related to the inhomogeneous distribution of director orientations. This strain is usually characterized by the director gradient or related micro-rotation. We think that in contrast, this common type of orientation elasticity in semi-flexible LC polymers is of secondary importance. The main idea of this paper is that the typical polymer elasticity related to changing lengths of macromolecular strands is the main reason for occurring viscoelastic effects in flow of these liquids. This type of elasticity is not represented in the popular Leslie-Ericksen theory [10] for small-molecule nematics. Even for the LC polymers of the first type (with rigid monomer units) we neglect here the effects of spatial gradient of director on viscoelastic response of this type of polymer nematics. This effect can be taken in account later, when developing more detailed theory for this type of LC polymers. We demonstrate that viscoelastic phenomena in LC polymers and elastomers associated with anisotropic relaxation times can be naturally described using formal approach of non-equilibrium thermodynamics. Our theory may also be applied to describe the behavior of concentrated suspensions of axially symmetric particles in polymer fluids.

**2. Uniaxial symmetry**

In liquids crystals, orientation phenomena occur along with common transport processes. Before developing irreversible thermodynamics of viscoelasticity in polymer LC's, it is worth to remind the description [25] of orientation for rigid or semi-flexible axially symmetric molecules that form nematics.

The alignment of molecules is characterized by their adjustable orientation and described by averaging properties of the unit vector $\underline{s}$ directed along the axis of molecular symmetry. The evolution behavior of this vector can be in general described by a non-Gaussian stochastic process, with infinite set of the moments $<s_i>$, $<s_i s_j>$ etc. Here the averaging is performed over the set of all realizations. Since $\underline{s}$ and $-\underline{s}$ vectors are equivalent for nonpolar molecules, the odd moments vanish. Thus the simplest statistical parameters that describe the molecular orientation are the second $<s_i s_j>$ and the fourth $<s_i s_j s_k s_l>$ moments. Therefore the second and fourth order traceless tensors

$$S_{ij} = \langle s_i s_j \rangle - 1/3 \delta_{ij} \qquad (1)$$

$$S_{ijkl} = \langle s_i s_j s_k s_l \rangle - 1/15(\delta_{ij}\delta_{kl} + \delta_{ik}\delta_{jl} + \delta_{il}\delta_{jk}) \qquad (2)$$

are defined so that they vanish when orientation is equiprobable. To describe the alignment in liquid crystals a second rank order tensor is commonly used. In uniaxial liquid crystals it takes the form

$$S_{ij} = S(n_i n_j - 1/3 \delta_{ij}). \qquad (3)$$

The *order tensor* (3) defines the unit vector $\underline{n}$, *director* that is directed along the material local symmetry axis. The individual molecular orientations are distributed around vector $\underline{n}$. It should be noted that $\underline{n}$ could not be identified with a simple averaging of the vector $\underline{s}$ because the odd moments of the molecular orientation vanish.

The scalar order parameter $S(t) (\in [0,1])$ called the *degree of alignment* is the macroscopic measure of alignment described by director. The function $S(t)$ typically decreases with increasing temperature. Thus the case $S = 0$ corresponds to the complete disorder, whereas $S = 1$ to the perfect order when all molecules are aligned in parallel. We further assume that the direction of preferred orientation changes only due to external actions. This leads to a simple and more tractable theory. In this case, parameter $S$ can be treated as a constant. Also, the second and higher order moments can also be expressed by the vector $\underline{n}$. Therefore instead of infinite set of tensor parameters the orientation can be described by a single unit vector $\underline{n}$. Strictly speaking, the uniaxial symmetry can be justified for well-ordered samples of nematic LC polymers when flow is slow enough. Nevertheless, we assume below that this type of symmetry holds even for strong flows.

### 3. Non-equilibrium thermodynamics

*3.1. Thermodynamic variables*

To develop a continuum theory of nematic LC polymers we introduce along with temperature $T$ two thermodynamic hidden variables. The first is a unit vector $\underline{n}$ director.

This vector characterizes the properties of uniaxial anisotropy in LC polymers. Additionally, such a hidden variable as the transient elastic strain tensor $\underline{c}$ (second rank, symmetric and positive definite) characterizing the elastic deformations in polymeric fluids [23,24] is also introduced as another thermodynamic variable.

In the common case of isotropic elastic liquids, the transient elastic strain tensor $\underline{c}$ can be related to the "conformation tensor", average dyadic $<\underline{RR}>$ built up on the end-to-end vector $\underline{R}$ between the ends of macromolecular strands, and normalized by the equilibrium value. The same sense of this tensor can be used for the flexible enough LC polymers and elastomers from the second group. In fact this measure of elastic strain or related to that, the tensor $\underline{c}^{1/2}$, has been used for describing equilibrium properties [13] of LC elastomers. Using the same approach to more rigid LC macromolecules from the first type of LC polymers, we can roughly consider the molecular sense of tensor $\underline{c}$ as proportional to the dyadic $<\delta\underline{R}\delta\underline{R}>$, normalized to its equilibrium value. Here $\delta\underline{R}$ is the variation of the end-to-end vector due to imposed stress field, such that $\underline{R} = \underline{R}^* + \delta\underline{R}$ and $<\underline{R}^*\delta\underline{R}>=0$, where the vector $\underline{R}^*$ is associated with the definition of director $\underline{n}_0$. Since defined partitioning of vector $\underline{R}$ into $\underline{R}^*$ and $\delta\underline{R}$ is vague, the above microscopic definitions of the above two hidden variables for LC polymers of this type is by no means precise and has only illustrative character.

Another hidden variable, which may play important role for very fast motions, is the kinematic ltensor (or corresponding vector) of internal rotations. This variable is presented in all the theories for LC elastomers [12,13] and commonly related to the occurring the non-symmetrica stress. It is possible to show, however, that without external (e.g. madnetic) fields and neglecting the director gradient and the inertia effects of internal rotations, the stress tensor must be symmetric. These conditions are also applied to the equilibrium theories for liquid crystalline nematic elastomers [12,13]. In this regard, the geometrical and kinematic characteristics of internal rotations can be easily calculated through the state variables and the velocity gradient of continuum. We adopt below all the basic assumptions of the non-equilibrium thermodynamics [26,27], such as "closeness" to the equilibrium (and therefore the existence of thermodynamic functions) and the local equilibrium assumption. We do not assume in this Section smallness of elastic strains and develop below a highly nonlinear theory, remaining however in the frame of the quasi-linear thermodynamic approach [23,24,28]. Although this approach is quite general, being mostly interested in fluid mechanics and rheology, we consider below only the flow effects.

*3.2. Free energy*

When the external (electrical or magnetic) fields are absent and the effects of director gradient and inertia of internal rotations are ignored, the stress tensor is symmetric and the full energy of system per mass unit consists of kinetic and internal energies. Being interested mostly in isothermal situations, we can use instead of internal energy, the Helmholtz free energy $f$ (per mass unit), which depends on temperature, $T$, and the above two thermodynamic parameters, $\underline{n}$ and $\underline{\underline{c}}$:

$$f = f(T, \underline{n}, \underline{\underline{c}}). \qquad (4)$$

Here all the anisotropic properties of the system are described by the vector $\underline{n}$. It means that the free energy $f$ depends on the three basic invariants $I_1, I_2, I_3$ of tensor $\underline{\underline{c}}$, and the two mixed invariants $I_1^a$ and $I_2^a$ of tensors $\underline{n}$ and $\underline{\underline{c}}$,

$$I_1 = tr\underline{\underline{c}}, \quad I_2 = 1/2(I_1^2 - tr\underline{\underline{c}}^2), \quad I_3 = \det\underline{\underline{c}}, \quad I_1^a = tr(\underline{nn} \cdot \underline{\underline{c}}), \quad I_2^a = tr(\underline{nn} \cdot \underline{\underline{c}}^{-1}). \qquad (5)$$

Due to eq.(5), it is seen that the free energy (4) presented as

$$f = f(T, I_1, I_2, I_3, I_1^a, I_2^a), \qquad (4a)$$

is invariant relative to the transformation $\underline{n} \to -\underline{n}$. In the case of incompressibility, additional constraint, $\det\underline{\underline{c}} = 1$, might be employed even in non-equilibrium case, as was argued in [23]. We assume that the general expression (4) satisfies the thermodynamic stability constraints. It means that $f$ is a convex function of thermodynamic variables $\underline{n}$ and $\underline{\underline{c}}$, and goes to zero at equilibrium. It should be noted that in incompressible case when $I_3 = 1$, a linear combination of the invariants (5) will serve as an anisotropic analog to the well-known Moony-Rivlin elastic potential in the isotropic case. Moreover, a linear combination of $I_1$ and $I_2^a$ which can be obtained from the Warner potential [13], is the anisotropic analog of classic entropic elastic potential for isotropic rubber elasticity.

*3. Dissipation*

Utilizing the laws of conservation, mass energy and momentum and using the routine derivation procedure [26-28], one can obtain the expression for the dissipation $D$ (or the entropy production $P_s$) in the system under isothermal condition:

$$D \equiv TP_s\big|_T = tr(\underline{\underline{\sigma}} \cdot \underline{\underline{e}}) - \rho \dot{f}\big|_T = tr(\underline{\underline{\sigma}} \cdot \underline{\underline{e}}) - \rho tr(\partial f/\partial \underline{\underline{c}} \cdot \underline{\underline{\dot{c}}}) - \rho \partial \tilde{f}/\partial \underline{n} \cdot \underline{\dot{n}} \quad (\geq 0). \qquad (6)$$

The symbol $\geq$ in the right-hand side of (6) means that according to the Second Law, the dissipation is positive for all irreversible processes and vanishes when the system approaches the thermodynamic equilibrium. In Eq.(6), $\underline{\underline{\sigma}}$ is the symmetric stress tensor, $\underline{\underline{e}}$ is the symmetric strain rate tensor, $\underline{\underline{e}} = 1/2[\nabla \underline{u} + (\nabla \underline{u})^T]$, where $\underline{u}$ is the velocity vector, and $\rho$ is the density. The overdots in eq.(6) mean the "material" differentiation operation, $d/dt = \partial/\partial t + \underline{u} \cdot \nabla$, in a moving continuum. To hold the constraint $|\underline{n}| = 1$ during the differentiation one should introduce the scalar Lagrange multiplier $q$ and modify the free energy $f$ for $\tilde{f}$ as: $\tilde{f} = f(T, \underline{n}, \underline{\underline{c}}) - 1/2 q \underline{n} \cdot \underline{n}$. The same is also applied when there is an incompressibility constraint.

*3.4. Choice of thermodynamic fluxes and forces*

In the general approach of non-equilibrium thermodynamics, the dissipation is typically represented as a bilinear form, $D = \sum_k X_k Y_k$, where variables $X_k$ and $Y_k$ that might have different tensor structures for different values of $k$, are called respectively thermodynamic forces and thermodynamic fluxes. In continuum mechanic applications the thermodynamic fluxes $Y_k$ are usually kinematic variables, and conjugated to them thermodynamic forces $X_k$ characterize dynamic variables. The stress and strain rate tensors, $\underline{\underline{\sigma}}$ and $\underline{\underline{e}}$, represent respectively the external conjugated thermodynamic force and flux.

Except the linear case, determining the thermodynamic forces and fluxes related to the hidden variables is not a trivial problem. The formal difficulty here roots in the ill posedness of establishing tensor variables from their scalar products presented in dissipation. Thus additional physical arguments, along with satisfaction of some limiting conditions, should be employed to single out the expressions for thermodynamic fluxes and forces for the hidden thermodynamic variables $\underline{\underline{c}}$ and $\underline{n}$.

The values $\underline{\dot{\underline{c}}}$ and $\underline{\dot{n}}$ in eq.(6) cannot serve as thermodynamic fluxes since they are not frame invariant. However using the invariance of such a scalar as dissipation in eq.(6) relative to the time dependent rotations of the actual (say, Cartesian) coordinate system, the full time differentiation operations in eq.(6) are evidently extended to the co-rotational or Jaumann time derivatives denoted below by over-circles. One can prove this assertion directly using eq.(4a). Thus instead of eq.(6) one can use the equation:

$$D = TP_s\big|_T == tr(\underline{\underline{\sigma}} \cdot \underline{\underline{e}}) - \rho tr(\partial f / \partial \underline{\underline{c}} \cdot \underline{\underline{\overset{0}{c}}}) - \rho \partial f / \partial \underline{n} \cdot \underline{\overset{0}{n}}. \tag{6a}$$

$$\underline{\underline{\overset{0}{c}}} = \underline{\underline{\dot{c}}} - \underline{\underline{c}} \cdot \underline{\underline{\omega}} + \underline{\underline{\omega}} \cdot \underline{\underline{c}}; \qquad \underline{\overset{0}{n}} = \underline{\dot{n}} - \underline{n} \cdot \underline{\underline{\omega}}.$$

Here $\underline{\underline{\overset{0}{c}}}$ and $\underline{\overset{0}{n}}$ are the co-rotational (Jaumann) time derivatives and $\underline{\underline{\omega}} = 1/2[\nabla \underline{u} - (\nabla \underline{u})^T]$ is the anti-symmetric vorticity tensor. Thus the thermodynamic flux and conjugated to it thermodynamic force related to the director, are respectively the vectors $\underline{\overset{0}{n}}$ and $\rho \partial f / \partial \underline{n}$.

The problem with determining the thermodynamic flux and conjugated to it thermodynamic force related to second rank deformation tensor $\underline{\underline{c}}$ has been previously analyzed in Refs.[23,24] for the case of isotropic finite viscoelastic deformations. The solution of this problem suggested in these papers is extended below to anisotropic case. It has been proposed [23,24] to define the thermodynamic flux $\underline{\underline{e}}_e$ as in the case of anisotropic elastic solids (see Appendix A), which is the equilibrium limit for the complex liquid under study. This approach matches reasonably well the local equilibrium assumption and provides the theory with a proper limit to the equilibrium case, when $\underline{\underline{c}} \to \underline{\underline{B}}$. Here $\underline{\underline{B}}$ is the Finger tensor for the total deformation of the liquid. This heuristic approach has been justified when using a specific viscoelastic kinematics (e.g. see [23,24]). According to the approach, in non-equilibrium case, the thermodynamic flux $\underline{\underline{e}}_e$ is defined through the tensor $\underline{\underline{c}}$ and its co-rotational time derivative $\overset{0}{\underline{\underline{c}}}$ as in equilibrium case, i.e. using the relation similar to eq.(A3) in Appendix A:

$$\underline{\underline{e}}_e = \mathbf{L}^{-1}(\underline{\underline{c}}):\overset{0}{\underline{\underline{c}}}, \quad \text{or} \quad \overset{0}{\underline{\underline{c}}} = \underline{\underline{c}} \cdot \underline{\underline{e}}_e + \underline{\underline{e}}_e \cdot \underline{\underline{c}} \equiv \mathbf{L}(\underline{\underline{c}}):\underline{\underline{e}}_e . \tag{7}$$

Here the operators (rank of four tensors) $\mathbf{L}(\underline{\underline{c}})$ and $\mathbf{L}^{-1}(\underline{\underline{c}})$ are defined in formulae (A2) and (A5). Eq.(7) shows that the thermodynamic flux associated with tensor $\underline{\underline{e}}_e$ is a well defined function of tensors $\underline{\underline{c}}$ and $\overset{0}{\underline{\underline{c}}}$ being a linear homogeneous function of tensor $\underline{\underline{e}}_e$. We now can easily establish conjugated to $\underline{\underline{e}}_e$ thermodynamic force, $\underline{\underline{\sigma}}_e$, which we call the thermodynamic stress. To do that the second term in the right-hand side of eq.(6a) is transformed with the use of eq.(7) as follows:

$$\rho tr(\partial f / \partial \underline{\underline{c}} \cdot \overset{0}{\underline{\underline{c}}}) = \rho tr[\partial f / \partial \underline{\underline{c}} \cdot (\mathbf{L}(\underline{\underline{c}}):\underline{\underline{e}}_e)] = \rho tr[(\mathbf{L}(\underline{\underline{c}}): \partial f / \partial \underline{\underline{c}}) \cdot \underline{\underline{e}}_e] \equiv \rho tr(\underline{\underline{\sigma}}_e \cdot \underline{\underline{e}}_e). \tag{8}$$

Here using Eq.(7) the thermodynamic stress is determined as:

$$\underline{\underline{\sigma}}_e = \rho \mathbf{L}(\underline{\underline{c}}): \partial f / \partial \underline{\underline{c}} \equiv \rho(\underline{\underline{c}} \cdot \partial f / \partial \underline{\underline{c}} + \partial f / \partial \underline{\underline{c}} \cdot \underline{\underline{c}}) . \tag{9}$$

Expression for thermodynamic stress (9) obtained by this derivation is a natural generalization of the Murnaghan formula, $\underline{\underline{\sigma}}_e = 2\rho \underline{\underline{c}} \cdot \partial f / \partial \underline{\underline{c}}$, well known in isotropic finite elasticity [29] ($\underline{\underline{c}} = \underline{\underline{B}}$) and isotropic finite viscoelasticity [23,24] when the free energy function $f$ depends only on temperature and the tensor $\underline{\underline{c}}$.

It is important to notice that the thermodynamic stress $\underline{\underline{\sigma}}_e$ and conjugated to it thermodynamic flux $\underline{\underline{e}}_e$ have the limit transitions to the values defined in equilibrium (generally anisotropic) elasticity: $\underline{\underline{\sigma}}_e \to \underline{\underline{\sigma}}$, $\underline{\underline{e}}_e \to \underline{\underline{e}}$ when $\underline{\underline{c}} \to \underline{\underline{B}}$.

*3.5. Equivalent expressions for dissipation*

Substituting eq.(8) into eq.(6a) yields:

$$D = TP_s\big|_T = tr(\underline{\underline{\sigma}} \cdot \underline{\underline{e}}) - tr(\underline{\underline{\sigma}}_e \cdot \underline{\underline{e}}_e) + \underline{N} \cdot \overset{0}{\underline{n}}, \quad \underline{N} = -\rho \partial f / \partial \underline{n} \tag{6b}$$

Here the conjugated thermodynamic force and flux, $\underline{\underline{\sigma}}_e$ and $\underline{\underline{e}}_e$ are defined in eqs.(9) and (7) respectively. The thermodynamic force $\underline{N}$ called orientation force, is related to director $\underline{n}$. Because the director $\underline{n}$ has the unit length $(n=1)$, the orientation force $\underline{N}$ is changes for

$$\underline{\tilde{N}} = \underline{N} - q\underline{n}. \tag{10}$$

Here $q$ is the Lagrange multiplier defined above in Section 3.3 to hold the constraint $\underline{n} \cdot \underline{n} = 1$. Because of this constraint $\underline{n} \cdot \underline{\dot{n}} = \underline{n} \cdot \overset{0}{\underline{n}} = 0,$ and one can replace the vector $\underline{\tilde{N}}$ in eq.(6b) without loss of generality, by the vector $\underline{N}$.

The thermodynamic stress tensor $\underline{\underline{\sigma}}_e$ and conjugated to it flux $\underline{\underline{e}}_e$ have natural limits in the anisotropic solid equilibrium. Therefore it is convenient to introduce, as has been done for isotropic viscoelasticity [23,24], the non-equilibrium thermodynamic force $\underline{\underline{\sigma}}_p$ and flux $\underline{\underline{e}}_p$ that vanish in equilibrium:

$$\underline{\underline{\sigma}}_p = \underline{\underline{\sigma}} - \underline{\underline{\sigma}}_e; \quad \underline{\underline{e}}_p = \underline{\underline{e}} - \underline{\underline{e}}_e. \tag{11}$$

Unlike the tensors $\underline{\underline{\sigma}}_e$ and $\underline{\underline{e}}_e$, the tensors $\underline{\underline{\sigma}}_p$ and $\underline{\underline{e}}_p$ in eq.(11) have yet to be determined from the set of constitutive equations.

Substituting eqs.(10) and (11) into (6b) yields the most convenient form for dissipation:

$$D = TP_s\big|_T = tr(\underline{\underline{\sigma}}_p \cdot \underline{\underline{e}}) + tr(\underline{\underline{\sigma}}_e \cdot \underline{\underline{e}}_p) + \underline{N} \cdot \overset{0}{\underline{n}} \ (\geq 0). \tag{12}$$

Eq. (12) clearly demonstrates the three sources of dissipation in the system: (i) deviation of stress tensor from thermodynamic one ($\underline{\underline{\sigma}}_p \equiv \underline{\underline{\sigma}} - \underline{\underline{\sigma}}_e \neq \underline{\underline{0}}$), (ii) deviation of strain rate tensor from the thermodynamic one ($\underline{\underline{e}}_p \equiv \underline{\underline{e}} - \underline{\underline{e}}_e \neq \underline{\underline{0}}$), and (iii) the existence of the orientation force ($N \neq 0$).

*3.6. Thermodynamic equilibrium*

Due to eq.(12) the thermodynamic equilibrium ($D = 0$) is characterized as:

$$\underline{\underline{\sigma}}_e \to \underline{\underline{\sigma}}, \quad \underline{\underline{e}}_e \to \underline{\underline{e}}, \quad \underline{\tilde{N}} \to \underline{0} \ (\underline{\underline{c}} \to \underline{\underline{B}}). \tag{13}$$

Since in equilibrium, $\widetilde{\underline{N}}_0 = \underline{N}_0 - q_0 \underline{n}_0 = \underline{0}$, the quantity $q_0$ is found as $q_0 = \underline{N}_0 \cdot \underline{n}_0$. Hereafter $\underline{n}_0$ stands for the equilibrium value of director, which sometimes can be generated using the fields. Thus in equilibrium eq.(10) yields:

$$\partial f / \partial \underline{n}_0 = (\underline{n}_0 \cdot \partial f / \partial \underline{n}_0) \underline{n}_0. \qquad (14)$$

Thus in the case of uniaxially anisotropic finite elasticity, when the anisotropicity is characterized by the rigid director $\underline{n}_0$, the explicit closed set of constitutive equations is of the form:

- The free energy $f = f(T, \underline{n}_0, \underline{\underline{B}})$ is a convex function of director $\underline{n}_0$ and the Finger tensor $\underline{\underline{B}}$.

- The stress-strain-director relations:

$$\underline{\underline{\sigma}} = \rho(\underline{\underline{B}} \cdot \partial f / \partial \underline{\underline{B}} + \partial f / \partial \underline{\underline{B}} \cdot \underline{\underline{B}}); \quad \rho = (\det \underline{\underline{B}})^{-1/2}; \quad \partial f / \partial \underline{n}_0 = (\underline{n}_0 \cdot \partial f / \partial \underline{n}_0) \underline{n}_0. \qquad (15)$$

We could not find the first simple formula in (15) in such a comprehensive review of general theory of finite elasticity as Ref.[17] (Sect. 84). Eq. (14), establishes an anisotropic dependence of director on the deformation tensor $\underline{\underline{B}}$.

*3.7. General constitutive relations*

In the quasi-linear theory developed below, the constitutive relations between thermodynamic forces and fluxes are linear but the phenomenological coefficients are considered as functions of state variables. Since the state variables might be of different tensor dimensionality, the phenomenological coefficients in general represent some tensors of different tensor dimensionalities. This creates the stress (or flow induced) anisotropy even in inherently isotropic at rest liquid. The Onsager's symmetry of phenomenological coefficients has been proved for both linear and quasi-linear cases [26-28]. This symmetry simplifies the structure of constitutive equations. The positiveness of dissipation, i.e. the Second Law of thermodynamics, imposes some additional constraints on the phenomenological coefficients. Unlike linear case, in general quasi-linear theory for viscoelastic nematics considered below, there is no reason to single out "scalar phenomena", i.e. partitioning all the second rank tensors in deviator and spherical components.

In accordance with Eq.(12) the thermodynamic forces are defined as $\underline{\underline{\sigma}}_p$, $\underline{\underline{\sigma}}_e$ and $\underline{N}$. Conjugated to them thermodynamic fluxes respectively are: $\underline{\underline{e}}$, $\underline{\underline{e}}_p$ and $\overset{0}{\underline{n}}$. The quasi-linear phenomenological relations between the thermodynamic fluxes and forces with account for Onsager symmetry are:

$$e_{ij} = \mathbf{A}^{11}_{ijkl}\sigma^p_{lk} + \mathbf{A}^{12}_{ijkl}\sigma^e_{lk} + \mathbf{b}^{13}_{ij,k}N_k$$

$$e^p_{ij} = \mathbf{A}^{12}_{ijkl}\sigma^p_{lk} + \mathbf{A}^{22}_{ijkl}\sigma^e_{lk} + \mathbf{b}^{23}_{ij,k}N_k \qquad (16)$$

$$\overset{0}{n}_k = \mathbf{b}^{13}_{ij,k}\sigma^p_{ji} + \mathbf{b}^{23}_{ij,k}\sigma^e_{ji} + \mathbf{a}_{ki}N_i.$$

Here all the kinetic coefficients, the tensors of various rank, represented in Eq.(16) by bold symbols are some functions of the thermodynamic state variables: temperature $T$, tensor $\underline{\underline{c}}$ and vector $\underline{n}$.

The rank of fourth kinetic tensors $\mathbf{A}^{st}_{ijkl}(s,t=1,2)$ are symmetric by the first two and second two pairs of indices and by the transposition of the first and second pairs of indices. The rank of three kinetic tensors, $\mathbf{b}^{s3}_{ij,k}(s=1,2)$, are symmetric by the first pairs of indices and when $n=1$, should hold the evident constraints: $\mathbf{b}^{s3}_{ij,k}n_k = 0$. The second rank kinetic tensor $\mathbf{a}_{ki}$ is symmetric and when $n=1$, should also hold the constraint: $\mathbf{a}_{ki}n_k = 0$. We avoid presenting the explicit structure of these tensors since they are described by very cumbersome formulae with many scalar functions depending on basic scalar invariants (5). A large simplification can be achieved when the kinetic coefficients $\mathbf{A}^s_{ijk}(s,t=1,2)$ and $\mathbf{b}^{s3}_{ij,k}(s=1,2)$ in eq.(16) are the only functions of temperature and director $\underline{n}$.

Substituting Eq.(16) into Eq.(12) yields the expression for dissipation as the quadratic form of the thermodynamic forces:

$$D = TP_s\big|_T =$$
$$\mathbf{A}^{11}_{ijkl}\sigma^p_{ij}\sigma^p_{lk} + 2\mathbf{A}^{12}_{ijkl}\sigma^p_{lij}\sigma^e_{lk} + \mathbf{A}^{22}_{ijkl}\sigma^e_{ij}\sigma^e_{lk} + 2\mathbf{b}^{13}_{ij,k}\sigma^p_{ij}N_k + 2\mathbf{b}^{23}_{ij,k}\sigma^e_{ij}N_k + \mathbf{a}_{ik}N_iN_k. \qquad (17)$$

The positive definiteness of dissipation $D$ imposes some additional constraints on the kinetic tensors, e.g. positive definiteness of the rank of fourth tensors $\mathbf{A}^{11}$, $\mathbf{A}^{22}$, and the second rank tensor $\mathbf{a}$, and also produces some additional constraints imposed on the combination of these tensors. All these constraints guarantee the existence of the constitutive relations, inverse to these presented in Eq.(16). These inverse relations represent the "dual" quasi-linear relations between the thermodynamic forces, $\underline{\underline{\sigma}}_p$, $\underline{\underline{\sigma}}_e$ and $\underline{N}$, and the thermodynamic fluxes, $\underline{\underline{e}}$, $\underline{\underline{e}}_p$ and $\overset{0}{\underline{n}}$:

$$\sigma_{ij}^p = \widetilde{\mathbf{A}}_{ijkl}^{11} e_{lk} + \widetilde{\mathbf{A}}_{ijkl}^{12} e_{lk}^p + \widetilde{\mathbf{b}}_{ij,k}^{13} \overset{0}{n}_k$$

$$\sigma_{ij}^e = \widetilde{\mathbf{A}}_{ijkl}^{12} e_{lk} + \widetilde{\mathbf{A}}_{ijkl}^{22} e_{lk}^p + \widetilde{\mathbf{b}}_{ij,k}^{23} \overset{0}{n}_k \qquad (16a)$$

$$N_k = \widetilde{\mathbf{b}}_{ij,k}^{13} e_{ji} + \widetilde{\mathbf{b}}_{ij,k}^{23} e_{ji}^p + \mathbf{a}_{ki} \overset{0}{n}_i.$$

Corresponding dual presentation of dissipation as a quadratic form of thermodynamic fluxes is of the form:

$$D = \widetilde{\mathbf{A}}_{ijkl}^{11} e_{ij} e_{lk} + 2\widetilde{\mathbf{A}}_{ijkl}^{12} e_{lij}^p e_{lk} + \widetilde{\mathbf{A}}_{ijkl}^{22} e_{ij}^p e_{lk}^p + 2\widetilde{\mathbf{b}}_{ij,k}^{13} e_{ij} \overset{0}{n}_k + 2\widetilde{\mathbf{b}}_{ij,k}^{23} e_{ij}^p \overset{0}{n}_k + \widetilde{\mathbf{a}}_{ik} \overset{0}{n}_i \overset{0}{n}_k. \quad (17a)$$

Here the same constraints as discussed above are imposed on the kinetic coefficients.

Consider now the second relation in eq.(11). Multiplying scalarly this equation by the rank of four tensor $\mathbf{L}(\underline{\underline{c}})$ defined in eq.(7) reduces this equation to the form:

$$\overset{0}{\underline{\underline{c}}} - \underline{\underline{c}} \cdot (\underline{\underline{e}} - \underline{\underline{e}}_p) - (\underline{\underline{e}} - \underline{\underline{e}}_p) \cdot \underline{\underline{c}} \equiv \overset{\nabla}{\underline{\underline{c}}} + \underline{\underline{c}} \cdot \underline{\underline{e}}_p + \underline{\underline{e}}_p \cdot \underline{\underline{c}} = \underline{\underline{0}}. \qquad (18)$$

Here the symbol $\overset{\nabla}{}$ denotes the upper convected tensor time derivative. Eq.(18) has been independently obtained in Refs.[7,8] from using the method shown above and an analysis a special viscoelastic kinematics. When irreversible strain rate tensor $\underline{\underline{e}}_p$ is expressed via other thermodynamic forces and fluxes as shown in Eq.(16), Eq.(18) represents the *evolution equation* for thermodynamic parameter $\underline{\underline{c}}$. The evolution equation for another thermodynamic parameter, director $\underline{n}$, is presented as third equation in the set (16). It should be noted that when external electric or magnetic fields are imposed, an additional vorticity tensor related to inelastic rotations might also appear in Eq.(18) [23,24].

Eqs.(16) and (18) along with the definitions of $\underline{\underline{\sigma}}_e$ and $\underline{\underline{\sigma}}_p$ given in eqs.(7) and (11) represent a closed set of anisotropic viscoelastic constitutive equations for polymeric LC.

Consider now some important particular cases of these constitutive equations. Apart from the equilibrium case considered earlier, there are two additional types of anisotropic viscoelastic behavior. They are described as follows.

(i) *The solid-like behavior of the Kelvin-Voight type*. It is described when the irreversible strain rate $\underline{\underline{e}}_p = \underline{\underline{0}}$ in Eq.(12). In this case, $\underline{\underline{c}} = \underline{\underline{B}}$, i.e. elastic deformation is equal to the total one. The irreversibility come here from two sources, the difference between the actual $\underline{\underline{\sigma}}$ and thermodynamic (equilibrium) stress $\underline{\underline{\sigma}}_e$ ($\underline{\underline{\sigma}}_p \neq \underline{\underline{0}}$) and the evolution of the director ($\underline{N} \neq \underline{0}$). The constitutive equations here are described by a limit case of Eq.(16) where

$$\mathbf{A}^{12} = \mathbf{A}^{22} = \mathbf{0}; \quad \mathbf{b}^{23} = \mathbf{0}. \tag{19}$$

(ii) T*he liquid-like behavior of the Maxwell type.* It is described when in Eq.(12) the irreversible stress $\underline{\underline{\sigma}}_p = \underline{\underline{0}}$. It means that in this case, the thermodynamic stress $\underline{\underline{\sigma}}_e$ defined by eq.(8) coincides with the actual stress $\underline{\underline{\sigma}}$. The irreversibility here also comes from two sources, the occurrence of irreversible deformations ($\underline{\underline{e}}_p \neq \underline{\underline{0}}$) and the evolution of the director ($\underline{\widetilde{N}} \neq \underline{0}$). The constitutive equations here are described by a limit case of Eq.(16) where

$$\mathbf{A} \to \infty, \quad \mathbf{A}^{12} = \mathbf{0}, \quad \mathbf{b}^1 = \mathbf{0}. \tag{19a}$$

To use for solving problems the general structure of the nonlinear thermodynamic theory developed in this Section, one needs to specify the free energy function $f$ and the kinetic coefficients of various tensor dimensionality. The choice of the function $f$ can be made relatively easy, because as mentioned, $f$ might be taken from the equilibrium theories (e.g. [13]). The problem of specifying the kinetic coefficient in the nonlinear approach is much more difficult to solve. For the weakly nonlinear theory, the choices of free energy and kinetic coefficients will be demonstrated below in Section 4.2. Some attempts to make this choice in the completely nonlinear theory will be made elsewhere.

## 4. Constitutive equations for Maxwell anisotropic viscoelastic liquids

*4.1. General nonlinear case*

The total closed set of constitutive equations for anisotropic viscoelastic liquids of Maxwell type is presented below since this case seems to be important for applications.

- Stable expression (4) for the free energy $f$:

$$f = f(T, \underline{n}, \underline{\underline{c}}). \tag{20.1}$$

- Expression (9) for the stress tensor $\underline{\underline{\sigma}}$:

$$\underline{\underline{\sigma}} = \rho(\underline{\underline{c}} \cdot \partial f / \partial \underline{\underline{c}} + \partial f / \partial \underline{\underline{c}} \cdot \underline{\underline{c}}). \tag{20.2}$$

- Expression in Eq.(6b) for the orientational force $\underline{N}$:

$$\underline{N} = -\rho \partial f / \partial \underline{n}. \tag{20.3}$$

- Evolution equation (18) for elastic strain tensor $\underline{\underline{c}}$:

$$\overset{0}{\underline{\underline{c}}} - \underline{\underline{c}} \cdot (\underline{\underline{e}} - \underline{\underline{e}}_p) - (\underline{\underline{e}} - \underline{\underline{e}}_p) \cdot \underline{\underline{c}} \equiv \overset{\triangledown}{\underline{\underline{c}}} + \underline{\underline{c}} \cdot \underline{\underline{e}}_p + \underline{\underline{e}}_p \cdot \underline{\underline{c}} = \underline{\underline{0}}. \tag{20.4}$$

- Quasi-linear phenomenological relations (16) with account for Eq.(19a):

$$e_{ij}^p = \mathbf{A}_{ijkl}\sigma_{lk} + \mathbf{b}_{ij,k} N_k \ ; \quad \overset{0}{n}_k = \mathbf{b}_{ij,k}\sigma_{ji} + \mathbf{a}_{ki} N_i. \tag{20.5}$$

Here the tensors $\mathbf{A}$, $\mathbf{b}$, and $\mathbf{a}$ of different ranks stand respectively for $\mathbf{A}^{22}$, $\mathbf{b}^{23}$, and $\mathbf{a}$ in Eq.(16). These tensors are some functions of tensor $\underline{\underline{c}}$ and director $\underline{n}$.

- In this case Eq. (17) for dissipation reduces to:

$$D = TP_s\big|_T = \mathbf{A}_{ijkl}\sigma_{ij}\sigma_{lk} + 2\mathbf{b}_{ij,k}\sigma_{ij} N_k + \mathbf{a}_{ik} N_i N_k \ (\geq 0). \tag{20.6}$$

- The conditions of positive definiteness of dissipation are:

$$\mathbf{A} > 0, \quad \mathbf{a} > 0, \quad \mathbf{a} - \mathbf{A}^{-1} : (\mathbf{b} : \mathbf{b}) > 0 \ (\mathbf{a}_{ij} - \mathbf{A}^{-1}_{rsmn}\mathbf{b}_{rs,i}\mathbf{b}_{mn,j} > 0). \tag{20.7}$$

- The constraints (20.7) also prove the existence of the inverse phenomenological relations (16a), which in our case have the form:

$$\sigma_{ij} = \widetilde{\mathbf{A}}_{ijkl} e_{lk}^p + \widetilde{\mathbf{b}}_{ij,k} \overset{0}{n}_k \ ; \quad N_k = \widetilde{\mathbf{b}}_{ij,k} e_{ji}^p + \mathbf{a}_{ki} \overset{0}{n}_i. \tag{20.5a}$$

- The corresponding expression for dissipation is:

$$D = TP\big|_{.} = \widetilde{\mathbf{A}}_{ijkl} e_{ij}^p e_{lk}^p + 2\widetilde{\mathbf{b}}_{ij,k} e_{ij}^p \overset{0}{n}_k + \widetilde{\mathbf{a}}_{ik} \overset{0}{n}_i \overset{0}{n}_k \geq 0. \tag{20.6a}$$

Equations (20.1)-(20.5(a)) represent the closed set of the constitutive equations for anisotropic viscoelastic liquids of Maxwell type.

*4.2. Weakly nonlinear viscoelastic nematodynamics for the Maxwell liquid*

When a characteristic relaxation time of the system is small enough, one can expect that the Weissenberg number of the system, $We = U\theta / L$, is also very small, even if the rates of deformations are high enough. Here $\theta$, $U$ and $L$ represent some characteristic relaxation time, velocity and length scale of flow. In this case, a simplification is possible, since the only constitutive non-linearity presented here will be due to the uniaxial anisotropy caused by the director. We will also consider here the common simplifying assumption of liquid incompressibility, $\rho = const$. In this case, the transient elastic strain tensor $\underline{\underline{\varepsilon}}$ can be considered as a small Henky tensor ($|\underline{\underline{\varepsilon}}| << 1$):

$$\underline{\underline{c}} = \underline{\underline{\delta}} + 2\underline{\underline{\varepsilon}} + O(\underline{\underline{\varepsilon}}^2). \tag{21}$$

Note that even in the weakly nonlinear case, we do not consider the transient elastic strain $\underline{\underline{\varepsilon}}$ to be infinitesimally small, i.e. $tr\underline{\underline{\varepsilon}} \neq 0$.

The free energy of the system vanishing in the equilibrium, ($\underline{\underline{\varepsilon}} \to 0, \underline{n} \to \underline{n}_0$) is represented with accuracy of $O(\underline{\underline{\varepsilon}}^2)$ as the positively definite quadratic form of tensor $\underline{\underline{\varepsilon}}$ and director $\underline{n}$,

$$\rho f \equiv w = 1/2 G_0 tr(\underline{\underline{\varepsilon}}^2) + G_1 \{tr(\underline{nn} \cdot \underline{\underline{\varepsilon}}^2) - [tr(\underline{nn} \cdot \underline{\underline{\varepsilon}})]^2\} + 1/2 G_2 tr(\underline{nn} \cdot \underline{\underline{\varepsilon}}). \tag{22}$$

Here $G_0$, $G_1$ and $G_2$ are temperature dependent shear moduli in the model. Eq.(22) holds $\underline{n} \to -\underline{n}$ invariance, convexity and vanishing at the equilibrium. The last term in Eq.(22) describes the yield in the system, with no vanishing stress at the equilibrium. The form of two first terms in free energy expression (22) accounts for the deviatoric component of stress tensor that will be demonstrated later.

Note that the first two terms in right-hand side of Eq.(22) are positively defined, as soon as the moduli $G_0$ and $G_1$ are positive. The stress tensor deviator (which is also denoted hereafter as) $\underline{\underline{\sigma}}$ is defined as:

$$\underline{\underline{\sigma}} \equiv \partial w / \partial \underline{\underline{\varepsilon}} = G_0 (\underline{\underline{\varepsilon}} - \underline{\underline{\delta}} tr \underline{\underline{\varepsilon}} / 3) + G_1 [\underline{nn} \cdot \underline{\underline{\varepsilon}} + \underline{\underline{\varepsilon}} \cdot \underline{nn} - 2\underline{nn} \cdot tr(\underline{nn} \cdot \underline{\underline{\varepsilon}})] + G_2 (\underline{nn} - 1/3\underline{\underline{\delta}}). \quad (23)$$

Eq.(23) where the last term in (3) represents the anisotropic yield shows that $tr\underline{\underline{\sigma}} = 0$..

The expression for the orientational force $\underline{N}$ with accuracy of $O(\underline{\varepsilon})$ is:

$$\underline{N} = -\partial w / \partial \underline{n} \approx G_2 \underline{\underline{\varepsilon}} \cdot \underline{n}. \quad (24)$$

The evolution equation for elastic strain is represented as:

$$\overset{\nabla}{\underline{\underline{\varepsilon}}} + \underline{\underline{e}}_p \approx \underline{\underline{e}} \quad (25)$$

Phenomenological relations are:

$$e^p_{ij} = \mathbf{A}_{ijkl}(\underline{n})\tilde{\sigma}_{lk} + \mathbf{b}_{ij,k}(\underline{n})N_k \; ; \quad \overset{0}{n}_k = \mathbf{b}_{ij,k}(\underline{n})\tilde{\sigma}_{ji} + \mathbf{a}_{ki}(\underline{n})N_i \; ; \quad \tilde{\sigma}_{ik} \equiv \sigma_{ik} - G_2(\underline{nn} - 1/3\underline{\underline{\delta}}) \quad (26)$$

The general expressions for the kinetic tensors $\mathbf{A}(\underline{n})$, $\mathbf{a}(\underline{n})$ and $\mathbf{b}(\underline{n})$ in Eq.(26) which take into account all the symmetry and traceless conditions are:

$$\mathbf{A}_{ijkl}(\underline{n}) = (1/\eta_0)\mathbf{\alpha}^{(1)}_{ijkl} + (1/\eta_1)\mathbf{\alpha}^{(2)}_{ijkl}(\underline{n}); \quad \mathbf{\alpha}^{(1)}_{ijkl} = 1/2(\delta_{ik}\delta_{jl} + \delta_{jk}\delta_{il} - 2/3\delta_{ij}\delta_{kl})$$
$$\mathbf{\alpha}^{(2)}_{ijkl}(\underline{n}) = 1/2(\delta^\perp_{ik}n_l n_j + \delta^\perp_{jk}n_l n_i + \delta^\perp_{il}n_k n_j + \delta^\perp_{jl}n_k n_i); \quad \delta^\perp_{ij} = \delta_{ij} - n_i n_j \quad (27)$$

$$\mathbf{a}_{ik} = (1/\eta_2)\delta^\perp_{ik}; \quad (28)$$

$$\mathbf{b}_{ij,k} = (1/\eta_3)\mathbf{\beta}_{ij,k}(\underline{n}); \quad \mathbf{\beta}_{ij,k}(\underline{n}) = \delta^\perp_{ik}n_j + \delta^\perp_{jk}n_i. \quad (29)$$

Here $\delta^\perp_{ij}$ is the transverse Kronecker symbol, which makes a projection on the direction orthogonal to the director; $\eta_k$ are temperature dependent "viscosities"; $\eta_1$ and $\eta_2$ being positive and $\eta_3$ sign indefinite. The properties of numerical tensors $\mathbf{\alpha}^{(1,2)}_{ijkl}(\underline{n})$ and $\mathbf{\beta}_{ij,k}(\underline{n})$ are considered in Appendix B.

Substituting eqs.(27)-(29) into eqs.(24) and (25) yields:

$$\overset{\nabla}{\underline{\underline{\varepsilon}}} + (1/\theta_0)(\underline{\underline{\varepsilon}} - \underline{\underline{\delta}}tr\underline{\underline{\varepsilon}}/3) + (1/\theta_1)[\underline{nn} \cdot \underline{\underline{\varepsilon}} + \underline{\underline{\varepsilon}} \cdot \underline{nn} - 2\underline{nn} \cdot tr(\underline{nn} \cdot \underline{\underline{\varepsilon}})] = \underline{\underline{e}} \quad (30)$$

$$\overset{0}{\underline{n}} = (1/\theta_2)[\underline{\underline{\varepsilon}} \cdot \underline{n} - \underline{n}tr(\underline{\underline{\varepsilon}} \cdot \underline{nn})]. \quad (31)$$

Here $\theta_0$, $\theta_1$ and $\theta_2$ are the relaxation times in the model, $\theta_0$ and $\theta_1$ being positive definite and $\theta_2$ the sign indefinite parameters. They are represented through the parameters $G_k$ and $\eta_k$ as follows:

$$\theta_0 = \frac{\eta_0}{G_0} \; ; \quad \theta_1 = \left( \frac{G_1}{\eta_0} + \frac{G_0 + G_1}{\eta_1} + \frac{G_2}{\eta_2} \right)^{-1} ; \quad \theta_2 = \left( \frac{G_0 + G_1}{\eta_3} + \frac{G_2}{\eta_2} \right)^{-1} . \tag{32}$$

The general initial conditions are:

$$\underline{\underline{\varepsilon}}\big|_{t=0} = \underline{\underline{\varepsilon}}_0, \qquad \underline{n}\big|_{t=0} = \underline{n}_0 . \tag{33}$$

Eqs. (23), (30) and (31) with the initial conditions (33) form a closed set of constitutive equations. These equations might be used for solving problems of polymer nematodynamics for the relatively rigid LC polymers.

4.3 *Relation to the Volkov's constitutive equations* [14-16]

We now consider linearized equations for orientation and constitutive relations when the transient elastic strain $\underline{\underline{\varepsilon}}$ is infinitesimal, i.e. $tr\underline{\underline{\varepsilon}} = 0$, and the yield in the system is ignored, i.e. $G_2 = 0$. In the linear case,

$$\underline{n} = \underline{n}_0 + \delta \underline{n} . \tag{34}$$

Here $\underline{n}_0$ is the director value in equilibrium, $\delta \underline{n}$ describes small rotations of director relative to its initial state. The vector $\delta \underline{n}$ is orthogonal to both $\underline{n}$ and $\underline{n}_0$ since they are unit vectors.

Multiplying Eq.(26) where in the linear case $\overset{0}{\underline{n}} = \delta \underline{\dot{n}} - \underline{n}_0 \cdot \underline{\underline{\omega}}$, by the third rank tensor $1/\eta_2 \beta_{mn,k}(\underline{n})$ and using Eqs.(B3) and (B4) from Appendix B yields with the accuracy of $O(\underline{\underline{\varepsilon}}^2)$ the linear equation for the director:

$$\tau \frac{d}{dt} \overset{0}{n}_k + \overset{0}{n}_k = \lambda (e_{kj} n_{0j} - n_{0k} n_{0i} n_{0j} e_{ij}); \quad \tau = \frac{\eta_0 \eta_1}{(G_0 + G_1)(\eta_0 + \eta_1)}, \quad \lambda = \frac{2\eta_0 \eta_1}{\eta_2 (\eta_0 + \eta_1)}. \tag{35}$$

Here $\lambda$ is the non-dimensional constant, and $\tau$ is the orientation relaxation time. Eq.(35) shows that the evolution equation for director $\underline{n}$ reduces to the Ericksen equation, when the characteristic time $\tau \ll L/U$. It means that in this case the Weissenberg number defined with relaxation time $\tau$ should be considerably small, i.e. $We_\tau = \tau U / L \ll 1$. In this limit, the Ericksen equation was used in paper [14] for viscoelastic case. Comparing that with other characteristic relaxation times that occur in Eq.(35), one can see a variety of asymptotic cases which can exist in flows. Using another approach, the paper [15] derived the orientation equation similar to Eq. (35), however in a less convenient form.

According to the eq. (35) for director, the small disturbance $\delta \underline{n}$ is of the order of velocity gradient. Using this fact along with evident simplifications: $\underline{\underline{N}} = O(\underline{\underline{\varepsilon}}^2) \approx 0$, $tr\underline{\underline{\varepsilon}} \approx 0$, $\overset{\nabla}{\underline{\underline{\varepsilon}}} \approx \dot{\underline{\underline{\varepsilon}}}$ in the linear case, and the formulae in Appendix B it is easy to represent the equations (23) and (30) as a anisotropic evolution (Maxwell-like) equation for the stress deviator:

$$\theta_{ijkl}(\underline{n}_0)\dot{\sigma}_{kl} + \sigma_{ij} \approx \eta_{ijkl}(\underline{n}_0)e_{kl} \tag{36}$$

Here, using Eqs.(B5), (B6) and (B13) from Appendix B, the anisotropic (the rank of four) tensors of viscosity and relaxation times are represented in the form:

$$\boldsymbol{\eta} = \eta_0[\boldsymbol{\alpha}^{(1)} - r_2\boldsymbol{\alpha}^{(2)}(\underline{n}_0)]; \qquad \boldsymbol{\theta} = (\eta_0/G_0)[\boldsymbol{\alpha}^{(1)} - r_1\boldsymbol{\alpha}^{(2)}(\underline{n}_0)] . \tag{37}$$

Here the numerical rank of fourth tensors $\boldsymbol{\alpha}^{(1)}(\underline{n}_0)$ and $\boldsymbol{\alpha}^{(2)}(\underline{n}_0)$ are defined in the eq.(27), and the non-dimensional coefficients $r_1$ and $r_2$ are:

$$r_1 = \frac{(G_0 + G_1)/\eta_1 + G_1/\eta_0}{(G_0 + G_1)(1/\eta_1 + 1/\eta_0)}; \qquad r_2 = \frac{\eta_0}{\eta_1 + \eta_0} . \tag{38}$$

The expression for dissipation (20.6), written with accuracy of $O(\underline{\underline{\varepsilon}}^2)$, is:

$$D = TP_s|_T \approx \mathbf{A}_{ijkl}(\underline{n})\sigma_{ij}\sigma_{lk} = 1/\eta_0 tr(\underline{\underline{\sigma}}^2) + 1/\eta_1\{tr(\underline{n}_0\underline{n}_0 \cdot \underline{\underline{\sigma}}^2) - [tr(\underline{n}_0\underline{n}_0 \cdot \underline{\underline{\sigma}})]^2\} . \tag{39}$$

**Appendix A: Expression for thermodynamic flux through the Finger strain tensor and its co-rotational time derivative in equilibrium elastic case.**

The procedure for finding the proper thermodynamic flux associated with the co-rotational tensor time derivative $\overset{0}{\underline{\underline{B}}}$ of Finger total strain tensor $\underline{\underline{B}}$ (symmetric and positive definite) immediately follows from the well known kinematic relation [23,24]:

$$\overset{\nabla}{\underline{\underline{B}}} \equiv \overset{0}{\underline{\underline{B}}} - \underline{\underline{B}} \cdot \underline{\underline{e}} - \underline{\underline{e}} \cdot \underline{\underline{B}} = \underline{\underline{0}}, \tag{A1}$$

Here $\overset{\nabla}{\underline{\underline{B}}}$ is the upper convected tensor time derivative of Finger strain tensor $\underline{\underline{B}}$ and $\underline{\underline{e}}$ is the strain rate tensor. Consider now the problem to find the tensor $\underline{\underline{e}}$ as a function of $\underline{\underline{B}}$ and $\overset{0}{\underline{\underline{B}}}$ from the equation:

$$\overset{0}{\underline{\underline{B}}} = \underline{\underline{B}} \cdot \underline{\underline{e}} + \underline{\underline{e}} \cdot \underline{\underline{B}} \equiv \mathbf{L}(\underline{\underline{B}}) : \underline{\underline{e}}; \quad L_{ijmn}(\underline{\underline{B}}) = \delta_{im} B_{jn} + B_{im} \delta_{jn}. \tag{A2}$$

Here the operator $\mathbf{L}(\underline{\underline{B}})$ is represented by a rank of fourth tensor whose Cartesian components $L_{ijmn}(\underline{\underline{B}})$ are symmetric by the first and second pairs of indices and by the transposition of the first and second pairs of indices. Also, the tensor $\mathbf{L}(\underline{\underline{B}})$ is positively definite, since for any symmetric second rank tensor $\underline{\underline{x}}$,

$$tr[\mathbf{L}(\underline{\underline{B}}) : \underline{\underline{xx}}] = 2 B_{ik} x_{kj} x_{ji} > 0.$$

Thus the inverse operation, $\mathbf{L}^{-1}(\underline{\underline{B}})$ does exists and is represented also by a rank of fourth tensor with the same property of symmetry as for the tensor $\mathbf{L}(\underline{\underline{B}})$. Therefore due to Eq.(A2) the strain rate tensor $\underline{\underline{e}}$ is expressed as a function of $\underline{\underline{B}}$ and $\overset{0}{\underline{\underline{B}}}$ as follows:

$$\underline{\underline{e}} = \mathbf{L}^{-1}(\underline{\underline{B}}) : \overset{0}{\underline{\underline{B}}}. \tag{A3}$$

The right-hand side of eq.(A3) is represented as a linear homogeneous function of $\overset{0}{\underline{\underline{B}}}$ and an isotropic tensor function of tensor $\underline{\underline{B}}$. The explicit presentation of tensor $\mathbf{L}^{-1}(\underline{\underline{B}}) : \overset{0}{\underline{\underline{B}}}$ is then of the form:

$$\mathbf{L}^{-1}(\underline{\underline{B}}) : \overset{0}{\underline{\underline{B}}} = \frac{1}{2}(I_1 I_2 - I_3)^{-1}[(I_1^2 + I_2)\overset{0}{\underline{\underline{B}}} + \underline{\underline{B}} \cdot \overset{0}{\underline{\underline{B}}} \cdot \underline{\underline{B}} + I_1 I_3 \underline{\underline{B}}^{-1} \cdot \overset{0}{\underline{\underline{B}}} \cdot \underline{\underline{B}}^{-1} -$$
$$- I_1(\underline{\underline{B}} \cdot \overset{0}{\underline{\underline{B}}} + \overset{0}{\underline{\underline{B}}} \cdot \underline{\underline{B}}) - I_3(\underline{\underline{B}}^{-1} \cdot \overset{0}{\underline{\underline{B}}} + \overset{0}{\underline{\underline{B}}} \cdot \underline{\underline{B}}^{-1})]. \tag{A4}$$

**Appendix B: Properties of constitutive tensors characterizing anisotropy in theory of weakly elastic nematic liquid crystals**

The non-dimensional kinetic tensors, $\boldsymbol{\alpha}^{(1)}_{ijkl}$, $\boldsymbol{\alpha}^{(2)}_{ijkl}$ and $\boldsymbol{\beta}_{ij,k}$, defined in eqs.(27)-(29) of the main text possess the following multiplicative properties that can be checked by direct calculations:

$$\boldsymbol{\alpha}^{(1)}_{ijmn}\boldsymbol{\alpha}^{(2)}_{nmkl}(\underline{n}) = \boldsymbol{\alpha}^{(2)}_{ijmn}(\underline{n})\boldsymbol{\alpha}^{(1)}_{nmkl} = \boldsymbol{\alpha}^{(2)}_{ijkl}(\underline{n}); \tag{B1}$$

$$\boldsymbol{\alpha}^{(1)}_{ijmn}\boldsymbol{\alpha}^{(1)}_{nmkl}(\underline{n}) = \boldsymbol{\alpha}^{(1)}_{ijkl}(\underline{n}); \quad \boldsymbol{\alpha}^{(2)}_{ijmn}\boldsymbol{\alpha}^{(2)}_{nmkl}(\underline{n}) = \boldsymbol{\alpha}^{(2)}_{ijkl}(\underline{n}); \tag{B2}$$

$$\boldsymbol{\beta}_{ij,k}\boldsymbol{\alpha}^{(1)}_{ijmn} = \boldsymbol{\beta}_{mn,k}; \tag{B3}$$

$$\boldsymbol{\beta}_{ij,k}\boldsymbol{\alpha}^{(2)}_{ijmn} = \boldsymbol{\beta}_{mn,k}. \tag{B4}$$

Consider now a rank of four tensor, $\mathbf{S}_{ijkl}(\underline{n})$, characterizing the uniaxial anisotropy:

$$\mathbf{S}_{ijkl}(\underline{n}) = S_0\boldsymbol{\alpha}^{(1)}_{ijkl} + S_1\boldsymbol{\alpha}^{(2)}_{ijkl}(\underline{n}); \quad S_0 > 0, \quad S_1 > 0. \tag{B5}$$

We assume that tensor $\mathbf{S}_{ijkl}(\underline{n})$ is positively definite, and finding the explicit expression for its inverse, i.e. the rank of four tensor $\mathbf{S}^{-1}_{ijkl}$, prove then the assumed positively definiteness. The existence of the tensor $\mathbf{S}^{-1}_{ijkl}$ means that any two symmetric traceless second rank tensors $\underline{\underline{x}}$ and $\underline{\underline{y}}$ can be related by the reciprocal equations,

$$\underline{\underline{y}}_{ij} = \mathbf{S}_{ijkl}(\underline{n})\underline{\underline{x}}_{kl}; \quad \underline{\underline{x}}_{ij} = \mathbf{S}^{-1}_{ijkl}(\underline{n})\underline{\underline{y}}_{kl}. \tag{B6}$$

We initially prove that the reciprocal tensor $\mathbf{a}^{-1}$ can be found from the equation:

$$\mathbf{S}_{ijkl}(\underline{n})\mathbf{S}^{-1}_{klmn}(\underline{n}) = \boldsymbol{\alpha}^{(1)}_{ijmn}. \tag{B7}$$

It is seen that the positively definite tensor $\boldsymbol{\alpha}^{(1)}_{ijmn}$ plays the role of unit tensor for traceless rank of four symmetric tensors. Due to the same reasons as those used for presentation of eq.(27), the general presentation for $\mathbf{S}^{-1}$ is of the same form as shown in Eq.(B5):

$$\mathbf{S}^{-1}_{ijkl}(\underline{n}) = \tilde{S}_0\boldsymbol{\alpha}^{(1)}_{ijkl} + \tilde{S}_1\boldsymbol{\alpha}^{(2)}_{ijkl}(\underline{n}). \tag{B.8}$$

Unknown constants $\tilde{S}_0$ and $\tilde{S}_1$ can be found by substituting Eqs.(B5) and (B8) into Eq.(B7) to find the solution of the problem as:

$$\mathbf{S}^{-1}_{ijkl}(\underline{n}) = \frac{1}{S_0}\left[\boldsymbol{\alpha}^{(1)}_{ijkl} - \frac{S_1}{S_0 + S_1}\boldsymbol{\alpha}^{(2)}_{ijkl}(\underline{n})\right] \quad . \tag{B9}$$

It is now easy to prove that the fourth rank tensor $\mathbf{S}^{-1}$ defined by eq.(B9) is positively definite. Indeed, due to Eq.(B9),

$$\forall \underline{\underline{x}} \quad \{tr\underline{\underline{x}} = 0, \underline{\underline{x}} = \underline{\underline{x}}^T\}: \qquad \underline{\underline{x}}:\mathbf{S}^{-1}:\underline{\underline{x}} \equiv \mathbf{S}^{-1}_{ijkl}(\underline{n})x_{ij}x_{kl} =$$

$$\frac{1}{S_0}\left[tr(\underline{\underline{x}}^2) - \frac{S_1}{S_0 + S_1}\{tr(\underline{\underline{x}}^2 \cdot \underline{nn}) - [tr(\underline{\underline{x}} \cdot \underline{nn})]^2\}\right] \geq \frac{S_0 tr(\underline{\underline{x}}^2) + S_1[tr(\underline{\underline{x}} \cdot \underline{nn})]^2}{S_0(S_0 + S_1)} > 0.$$

Here the evident inequality, $tr(\underline{\underline{x}}^2 \cdot \underline{nn}) \leq tr(\underline{\underline{x}}^2)$, has been used. The above inequality also proves that tensor $\mathbf{S}(\underline{n})$ defined in Eq.(B5) is positively defined.

**CONCLUSIONS**

A thermodynamically based nonlinear continuum theory was developed to describe nematodynamics of a viscoelastic liquid. The viscoelastic behavior of the liquid is described by two hidden variables, elastic strain tensor and a vector (director), characterizing anisotropy. Assuming that this approach is applicable to describe anisotropic viscoelastic properties in flows of semi-flexible polymer nematic LC's, a qualitative molecular interpretation of the hidden variables were proposed

In the present simplified theory, designed to describe the rheology and fluid mechanics of polymer semi-flexible nematics, effect of director gradient is neglected. This is in marked contrast to the common theoretical approach to low molecular nematic LC's. However, including the director gradient in the number of hidden thermodynamic variables might be necessary for more general description of polymer nematics under external electrical or magnetic fields. When these fields are absent, the stress tensor can be treated as symmetrical. Then the theory describes nonlinear anisotropic viscoelasticity (anisoropic viscosity and relaxation) of polymer nematics along with evolution equation for director in flows of liquid crystalline polymers. For weakly nonlinear case with infinitesimal transient elastic tensor the anisotropic behavior is described by few temperature dependent parameters. In this case the evolution equation for director reminds the Ericksen's equation, but with an additional relaxation term.

It should be finally noted that the proposed theory could also be applicable to the description of relatively slow flows of anisotropic suspensions of nematic type (e.g. with particles that are either prolate or oblate ellipsoids of revolution) in a viscoelastic fluid. To describe more rapid flows one should include into consideration also the inertial effects of internal rotations.

**References**

[1] E.T. Samulski, Physics Today, May 1982, 40.

[2] P.G. de Gennes, *The physics of liquid crystals*, Oxford Press, New York, 1974.

[3] M. Doi, J. Pol. Sci.: Pol. Phys. Ed. (1981) 229.

[4] M. Doi and S.F. Edwards, *The Theory of Polymer Dynamics*, Clarendon Press, Oxford (1986), Chs.8-10

[5] G. Marrucci, F. Greco, Adv. Chem. Phys., 53 (1993) 331.

[6] R.G. Larson, *The Structure and Rheology of Complex Fluids*, Oxford University Press, Oxford, 1998.

[7] J.J. Feng, G. Sgalari, and L.G. Leal, J. Rheol. 44 (2000) 1085

[8] B.J. Edwards, A.N. Beris, M. Grmela, Mol. Liq. Cryst, 201 (1991) 51

[9] A.N. Beris and B.J. Edwards, *Thermodynamics of Flowing Systems*, Oxford University Press, Oxford, 1999

[10] I.E. Dzyaloshinskii and G.E. Volovick, Ann. Phys., 125 (1980) 67

[11] L.G. Larson and D.W. Mead, J. Rheol. 33 (1989) 185

[12] de Gennes, in: W. Helfrich, G. Kleppke (Eds), *Liquid Crystals in One- and Two Dimensional Order,* Springer, Berlin (1980) 231

[13] M. Warner, Mech. Phys. Solids, 47 (1999) 1355

[14] V.S. Volkov and V.G. Kulichikhin, J. Rheol. 34 (1990) 281

[15] V.S. Volkov, in: Emri I (ed.) Progress and Trends in Rheology V (Proc. of the Fifth European Rheology Conference, 240-241, Ljubljana, Slovenia, 1998).

[16] V.S. Volkov, V.G. Kulichikhin, Rheol. Acta, 39, (2000) 360.

[17] R.S. Porter and J.F. Johnson, in *Rheology*, Vol.4, F.R. Eirich, Ed., Academic Press, New York, 1967.

[18] H. Pleiner and H.R. Brand, Mol. Cryst. Liq. Cryst. 199 (1991) 407

[19] H. Pleiner and H.R. Brand, Macromolecules, 25 (1992) 895

[20] A.D. Rey, Rheol. Acta., 34 (1995) 119.

[21] A.D. Rey, J. Non-Newt. Fluid. Mech. 58 (1995) 131

[22] D. Long and D.C. Morse, J. Rheol., 46 (2002) 49

[23] A.I. Leonov, A.N. Prokunin, *Nonlinear Phenomena in Flows of Viscoelastic Polymer Fluids*, Chapman & Hall, New York, 1994.

[24] A.I. Leonov, in: *Advances in the Flow and Rheology of Non-Newtonian Fluids,* D.A. Siginer, D. De Kee and R.P. Chhabra Eds. , Elsevier, New York,1999, 519.

[25] J. L. Ericksen, in: *Orienting Polymers,* J.L. Ericksen ed., Lecture Notes in Mathematics (Series) 1063, Springer, New York, 1984, 27.

[26] I. Prigogine, *Etude Thermodynamique des Phenomenes Irreversible*, Liege (1947).

[27] S.R. de Groot, P. Mazur, *Non-Equilibrium Thermodynamics*, North-Holland, Amsterdam (1962).

[28] I.Gyarmati, *Non-Equilibrium Thermodynamics*, Field Theory and Variational Principles, Springer, New York (1970)

[29] C. Truesdell and W. Noll, *The Non-Linear Field Theories of Mechanics*, Second. Ed., Springer, New York (1992).